\documentclass[oldversion, printer]{aa}
\usepackage[T1]{fontenc}
\usepackage{graphicx}
\usepackage{array}
\usepackage{subfigure}
\usepackage{txfonts}
\usepackage{natbib}
\usepackage{ulem}
\bibpunct{(}{)}{;}{a}{}{,}

\newcommand{\der}[2]{\ensuremath{\frac{{\rm d} #1}{{\rm d} #2}}}

\newcommand{\pder}[2]{\ensuremath{\frac{\partial #1}{\partial #2}}}

\newcommand{\be}{\begin{equation}}
\newcommand{\ee}{\end{equation}}
\newcommand{\bea}{\begin{eqnarray}}
\newcommand{\eea}{\end{eqnarray}}

\def\spose#1{\hbox to 0pt{#1\hss}}
\def\lta{\mathrel{\spose{\lower 3pt\hbox{$\mathchar"218$}}
        \raise 2.0pt\hbox{$\mathchar"13C$}}}
\def\gta{\mathrel{\spose{\lower 3pt\hbox{$\mathchar"218$}}
        \raise 2.0pt\hbox{$\mathchar"13E$}}}

\begin{document}
    \title{
    Leaving the ISCO: the inner edge of a black-hole accretion disk
    \\at various luminosities
    }
   \author{
          Marek A. Abramowicz \inst{1, 2, 5, 7}
          \and
              Micha{\l} Jaroszy{\'n}ski \inst{3}
          \and
              Shoji Kato \inst{4}
          \and
             Jean-Pierre Lasota\inst{5, 6}
          \and
              \\ Agata R{\'o}{\.z}a{\'n}ska\inst{2}
          \and
              Aleksander S\k{a}dowski \inst{2}
          }
   \institute{
             Department of Physics, G\"oteborg University,
             SE-412-96 G\"oteborg, Sweden    \\
             \email{Marek.Abramowicz@physics.gu.se}
         \and
             N. Copernicus Astronomical Center, Polish Academy
             of Sciences,
             Bartycka 18, PL-00-716 Warszawa, Poland \\
             \email{as@camk.edu.pl}, ~\email{agata@camk.edu.pl}
         \and
             Warsaw University Observatory,
             Al. Ujazdowskie 4, PL-00-478 Warszawa, Poland
             \\
             \email{mj@astrouw.edu.pl}
         \and
            2-2-2 Shikanodai-Nishi, Ikoma-shi, Nara 630-0114, Japan\\
            \email{kato.shoji@gmail.com}, \email{kato@kusastro.kyoto-u.ac.jp}
         \and
             Institut d'Astrophysique de Paris, UMR 7095 CNRS, UPMC Univ Paris 06, 98bis Bd Arago, 75014 Paris, France\\
             \email{lasota@iap.fr}
          \and
             Jagiellonian University Observatory, ul. Orla 171,
             PL-30-244 Krak{\'o}w,
             Poland
          \and
             Institute of Physics, Faculty of Philosophy and Science,
             Silesian University in Opava,
             Bezru{\v c}ovo n{\'a}m. 13, 746-01 Opava, Czech Republic
            }
\date{Received ????; accepted ???? }
  \abstract{The ``radiation inner edge'' of an accretion disk is
  defined as the inner boundary of the region from which most of
  the luminosity emerges. Similarly, the ``reflection edge'' is the
  smallest radius capable of producing a significant X-ray reflection
  of the fluorescent iron line. For black hole accretion disks
  with very sub-Eddington luminosities these and all other ``inner
  edges'' locate at ISCO. Thus, in this case, one may rightly
  consider ISCO as the unique inner edge of the black hole
  accretion disk. However, even for moderate luminosities,
  there is no such unique inner
  edge as differently defined edges locate at different
  places. Several of them are significantly closer to the black
  hole than ISCO. The differences grow with the increasing
  luminosity. For nearly Eddington luminosities, they
  are so huge that the notion of the inner edge losses all practical
  significance.}
\authorrunning{M.A. Abramowicz, M. Jaroszy{\'n}ski, S. Kato,
               J.-P. Lasota, A. R{\'o}{\.z}a{\'n}ska
               and A. S{\k a}dowski}
\titlerunning{Leaving the ISCO}
  \maketitle

%
\section{Introduction}
\label{section-introduction}
%

Accretion flows on to black holes must change character before
matter crosses the event horizon. Two reasons account for this
fundamental property of such flows. First, matter must cross the
black-hole surface at the speed of light as measured by a local
inertial observer \citep[see e.g.][]{gj-06}, so that if the flow
is sub-sonic far away from the black-hole (in practice it is
always the case) it will have to cross the sound barrier (well)
before reaching the horizon. This is the property of all realistic
flows independent of their angular momentum. The sonic surface in
question can be considered as the inner edge of the accretion
flow.

The second reason is related to angular momentum. Far from the
hole many (most probably most) rotating accretion flows adapt the
Keplerian angular momentum profile. Because of the existence of the
Inner-Most Stable Circular Orbit (ISCO) such flow must stop to be
Keplerian there. At high accretion rates when pressure gradients
become important the flow may extend below the ISCO but the
presence of the Inner-Most Bound Circular Orbit (IBCO) defines
another limit for a circular flow (the absolute limit being given
by the Circular Photon Orbit; the CPO). These critical circular
orbits provide another possible definition of the inner edge of the
flow, in this case of an accretion disk.

The question is: what is the relation between the accretion flow
edges? In the case of geometrically thin disks the sonic and
Keplerian edges coincide and one can define the ISCO as the inner
edge of such disks. \cite{paczynski-2000} showed rigorously that,
independent of viscosity mechanism, presence of magnetic fields
etc. the ISCO is the universal inner disk's edge for not too-high
viscosities. The case of thin disks is therefore
settled\footnote{In a recent paper \cite{Pennaetal-10} studied the
effects of magnetic fields on thin accretion disk (the disk
thickness $H/r \lta 0.07$, which corresponds to $L \lta
0.2\,L_{\rm Edd}$). They found that to within a few percent the
magnetized disks are consistent with the \cite{nov-tho-1973}
model, in which the inner edge coincides with the ISCO.}.

However, this is not the case of non-thin accretion disks, i.e.
the case of medium and high luminosities. The problem of defining
the inner edge of an accretion disk is not just a formal exercise.
\cite{afshordi-2003} explored several reasons which made
discussing the location of inner edge $r = r_{\rm in}$ of the
black hole accretion disks an interesting and important issue. One
of them was,
\par \hangindent=0.5cm {\it Theory of accretion disks is several
decades old. With time ever more sophisticated and more diverse
models of accretion onto black holes have been introduced.
However, when it comes to modeling disk spectra, conventional
steady state, geometrically thin-disk models are still used,
adopting the classical ``no torque'' inner boundary condition at
the marginally stable orbit.}

\noindent The best illustration of this fact is the case of the
state-of-art works on measuring the black hole spin $a$ in the
microquasar GRS 1915+105 by fitting its observed ``thermal state''
spectra to these calculated \citep[e.g.][]{sha-2008, mid-2009}.
These works use general relativistic version of the classical
Shakura-Sunyaev thin accretion disk model worked out by
\cite{nov-tho-1973}. The Novikov-Thorne model assumes that the
inner edge of the the disk $r_{\rm in} \equiv r_{\rm ISCO}$ is
also the innermost boundary of the radiating region.

Because the black hole mass of GRS 1915+105 is known and therefore
fixed ($M_0 = 14\,M_{\odot}\pm 4\,M_{\odot}$), the surface area
$A$ of the radiating region, calculated in the model, depends only
on the black hole unknown spin, $a^*$ ($a^* = Jc/GM^2$ with $J$
being the total
angular momentum of the black hole). In the
thermal state, the disk spectrum is close to that of a sum of
black body contributions from different radial locations. Its
shape is determined by the radial distribution of temperature,
which in the Novikov-Thorne model depends on the spin, $T = T(r,
a^*)$. The total radiation power $L$ is determined by the
``averaged'' temperature $T_0 = T_0(a^*)$ and the surface area $A
= A(a^*)$ of the radiating region, $L = \sigma T_0^4 A$. By
calculating the spectral shape and power for different $a^*$ in
the Novikov-Thorne model, one may find the best-fit estimates for
the spin-dependent temperature and area. This is just the main
idea of the spin estimate; details of the fitting are far more
complex \citep[see][]{sha-2008, davi-2005, str-2010} and include,
for example, a heuristic way of treating a contribution of
scattering in accretion disks atmosphere (i.e.  the ``hardening
factor'').

Results obtained this way by \cite{sha-2008} for GRS 1915+105
showed that $a^* = 0.99$ for the whole luminosity range $L <
0.2\,L_{\rm Edd}$. However, for $L > 0.2\,L_{\rm Edd}$, the spin
estimated by \cite{sha-2008} was much lower, $a^* \approx 0.8$.
The inconsistent spin estimates at different luminosities indicate
that some assumptions adopted by the Novikov-Thorne model are
wrong at high luminosities.

This is not a surprise, because there are several physical effects
known to be important at high luminosities, but ignored in the
classical Shakura-Sunyaev and Novikov-Thorne {\it thin} accretion
disk models. These effects are properly included in the {\it
slim}\footnote{The names {\it thin} and {\it slim} refer to the
dimensionless vertical geometrical thickness, $h = H/r$. For thin
disks it must be $h \ll 1$, while for slim disks a weaker
condition $h < 1$ holds.} accretion disks models, introduced by
\cite{abr-1988}. Advection is perhaps the best known of these
``slim disk effects'', but in the present context equally
important is a significant stress due to the radial pressure
gradient (for thin disks $dP/dr \approx 0$). The stress firmly
holds matter well inside ISCO and as a result of this, at high
luminosities the edge of the plunge-in region may be considerably
closer to the black hole than ISCO\footnote{Matter may be hold
well inside ISCO also by magnetic stresses, as pointed out by many
authors; see e.g. a semi-analytic model by
\cite{narayan-mad-2003}, or MHD numerical simulations by
\cite{noble-2010}, and references quoted in these papers.}.
%
%
\begin{figure}[h]
\centering
\includegraphics[width=9cm]{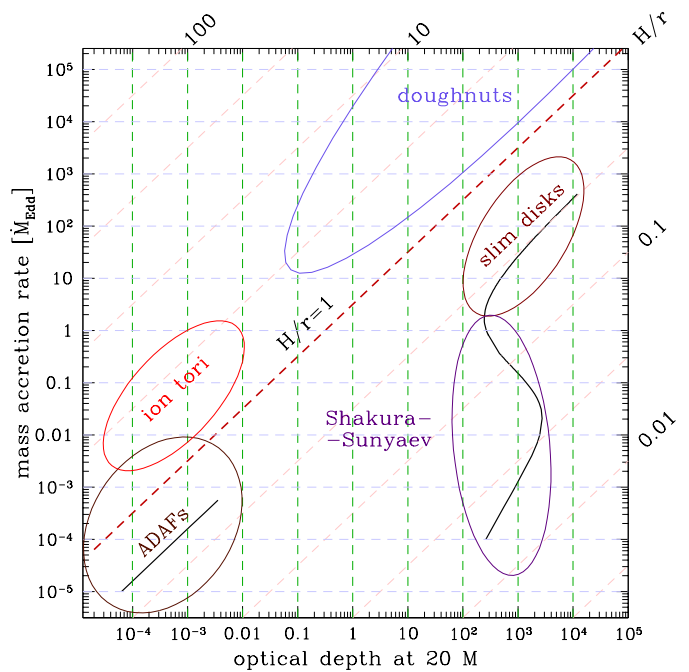}
\caption{The Figure illustrates a few best-known analytic and
semi-analytic solutions of the stationary black hole accretion
disks. Their location in the parameter space approximately
correspond to viscosity $\alpha = 0.1$ and radius $r = 20~{\rm M}$. For
detailed reviews of these solutions see, {\tt \tiny
http://www.scholarpedia.org/article/Accretion\_discs} or
\cite{kat-2008}.}
\label{fig:branches}
\end{figure}

Slim disks are assumed to be stationary and axially symmetric.
They are described by vertically integrated Navier-Stokes
hydrodynamical equations; no magnetic fields are considered. The
effective viscosity, believed to be generated by the MHD
turbulence \citep{balb-1991} is described by the ``$\alpha P$''
Shakura-Sunyaev ansatz. Figure~\ref{fig:branches} shows the slim
disk location with respect to other analytic and semi-analytic
disk models, in the parameter space $[\tau, h, {\dot m}]$
described by the vertical optical depth $\tau$, dimensionless
vertical thickness $h = H/r$ and dimensionless accretion rate
${\dot m}= {\dot M}/{\dot M}_{\rm Edd}$, where ${\dot M}_{\rm Edd}
= 16L_{\rm Edd}/c^2$ is the critical accretion rate approximately
corresponding to the Eddington luminosity ($L_{\rm Edd} =
10^{38}M/M_{\odot}\,$erg/s) in case of a disk around a
non-rotating black hole\footnote{Two warnings about notation. (i)
Many authors use a different definition, ${\dot M}_{\rm Edd} =
L_{\rm Edd}/c^2$. (ii) We often use the $c = 1 = G$ convention in
which $M = r_G = GM/c^2$.}.

In this paper, we discuss properties of the inner edge of slim
accretion disks around rotating black holes, using models similar
to those
calculated recently by \cite{sad-2009}\footnote{At {\tt \tiny
http://users.camk.edu.pl/as/slimdisks} a very detailed data
base for these solutions is given. It covers the whole parameter
space relevant for microquasars and AGN.}. For convenience, we
shortly remind the slim disk basic equations in the Appendix~A. In
the following Section \ref{section-definitions}, we list six possible
definitions of the inner edge. These definitions reflect different
(but partially overlapping) physical meanings and different
practical astrophysical applications. In the following six
Sections \ref{section-potential-spout}-\ref{section-reflection} we calculate
the slim disk locations of these six inner edges, and discuss
their astrophysical relevance. Some of the results presented here
have been anticipated previously by us and other authors in a
different context of the Polish doughnuts \citep[i.e. thick
accretion disks; see e.g. a short review by][]{paczynski-1998};
see also \cite{paczynski-2000} and \cite{afshordi-2003}.

%
%
\section{Definitions of the inner edge}
\label{section-definitions}

\citet{kro-2002} proposed several ``empirical'' definitions of the
inner edge, each serving a different practical purpose \citep[see
also the follow-up by][]{bec-2008}. We add to these a few more
definitions. The list of the inner edges considered in this paper
consists of\footnote{Krolik \& Hawley defined [4], [5], [6] above
and in addition [7], {\it the turbulence edge}, where
flux-freezing becomes more important than turbulence in
determining the magnetic field structure. Magnetic fields are not
considered for slim accretion disks, and we will not discuss
[7].},

\noindent [1] {\it The potential spout edge}~$r_{\rm in} = r_{\rm
pot}$, where the effective potential forms a self-crossing Roche
lobe, and accretion is governed by the Roche lobe overflow.

\noindent [2] {\it The sonic edge}~$r_{\rm in} = r_{\rm son}$,
where the transition from subsonic to transonic accretion occurs.
Hydrodynamical disturbances do not propagate upstream a supersonic
flow, and therefore the subsonic part of the flow is ``causally''
disconnected from the supersonic part.

\noindent [3] {\it The variability edge}~$r_{\rm in} = r_{\rm
var}$, the smallest radius where orbital motion of coherent spots
may produce quasi periodic variability.

\noindent [4] {\it The stress edge}~$r_{\rm in} = r_{\rm str}$,
the outermost radius where the Reynolds stress is small, and
plunging matter has no dynamical contact with the outer accretion
flow;

\noindent [5] {\it The radiation edge}~$r_{\rm in} = r_{\rm rad}$,
the innermost place from which significant luminosity emerges.

\noindent [6] {\it The reflection edge}~$r_{\rm in} = r_{\rm
ref}$, the smallest radius capable of producing significant
fluorescent iron line.

In the next six Sections we discuss the six edges one by one.

%
%
%
\section{The potential spout edge}
\label{section-potential-spout}
%

\begin{figure*}[t]
\centering
\includegraphics[width=1.\textwidth]{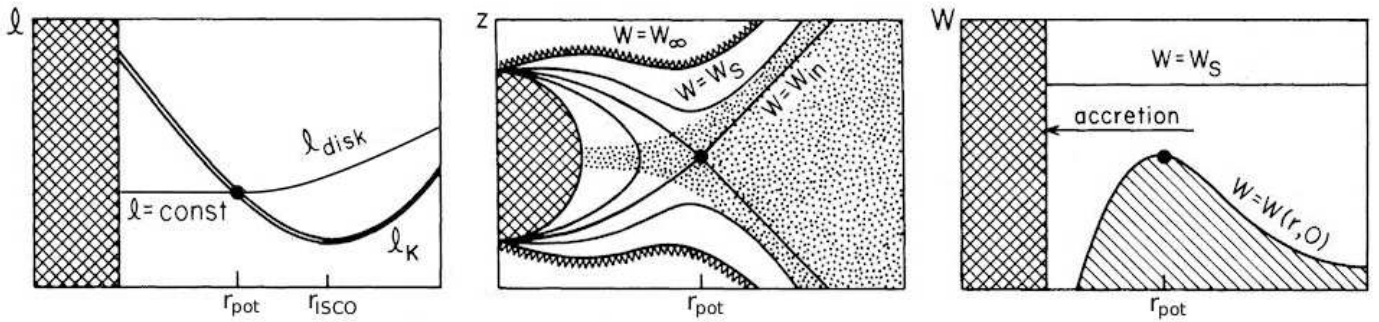}
\caption{An illustrative visualisation of the Roche lobe overflow.
The leftmost panel schematically presents disk angular momentum
profile and its relation to the Keplerian distribution. The middle
panel shows the equipotential surfaces. The dotted region denotes
the volume filled with accreting fluid. The rightmost panel
presents the potential barrier at the equatorial plane ($z=0$) and
the potential of the fluid ($W_S$) overflowing the barrier. The
figure is taken from {\tt \tiny
http://www.scholarpedia.org/article/Accretion\_discs}.
  }
\label{figure-roche-lobe-ill}
\end{figure*}

\begin{figure}[h]
\centering
\includegraphics[width=.5\textwidth]{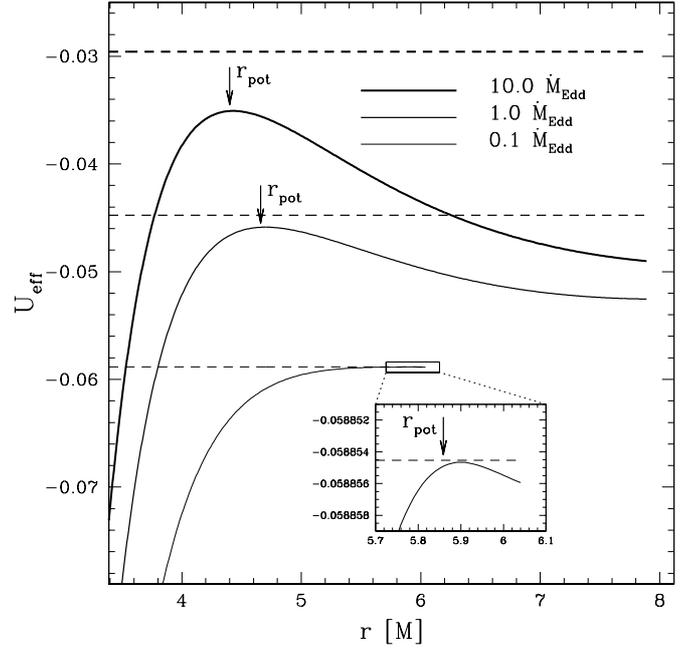}
\caption{Profiles of the effective potential near the potential
barrier (solid lines) for different accretion rates, $\alpha=0.01$
and $a^*=0$. The arrows indicate locations of the {\it inner edge
potential spout} - $r_{\rm pot}$ corresponding to the top of the
potential barrier. The horizontal dashed lines present the energy
of the gas overflowing the barrier calculated at $r_{\rm pot}$.
  }
\label{figure-roche-lobe}
\end{figure}

\begin{figure}[t]
\centering
\includegraphics[width=9.5cm]{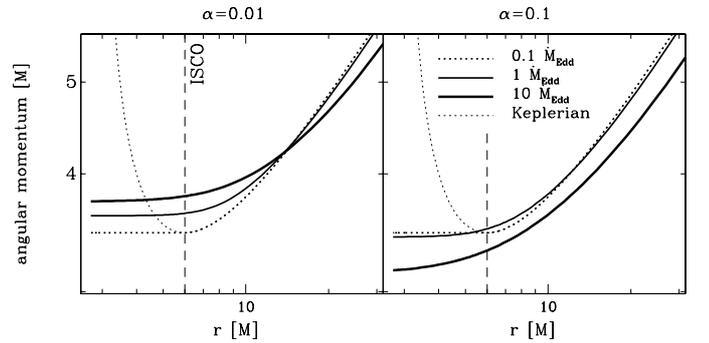}
\caption{Angular momentum profiles for slim disk solutions with
$\alpha=0.01$ (left panel) and $\alpha=0.1$ (right panel). In both
panels three curves are presented for sub-Eddingtonian,
Eddingtonian and super-Eddingtonian accretion rates. The thin dotted
line presents the Keplerian angular momentum profile. }
\label{figure-angular-momentum}
\end{figure}

The idea of the ``relativistic Roche lobe overflow'' governing accretion
close to the black hole was first explained by Paczy{\'n}ski
\citep[see][]{koz-1978}. Later, it was explored in detail by many
authors analytically \citep[e.g.][]{abr-1981, abr-1985} and by
large-scale hydrodynamical simulations \citep[e.g.][]{igu-1997}.
It became a standard concept in the black hole accretion theory.
Figure \ref{figure-roche-lobe-ill} schematically illustrates
the Roche lobe overflow mechanism.
The left-most panel presents demonstrative profile of disk angular momentum
which reaches the Keplerian value at the radius corresponding to
the self-crossing of the equipotential surfaces presented in the
middle panel. To flow through this ``cusp'' matter must have potential
energy higher than the value of the potential at this point - such
``potential barrier'' is crossed only when the matter overflows
its Roche lobe. Precise profiles of the
potential barriers and the angular momentum, calculated with the slim disk model, are presented
in Figs.~\ref{figure-roche-lobe} and \ref{figure-angular-momentum} ,
respectively.

%
%

%
%

The potential difference between the horizon and the spout is
infinite, and therefore no stress may prevent the matter located
there from plunging into the black hole. For radii greater that
$r_{\rm pot}$, the potential barrier at $r = r_{\rm pot}$ holds
the matter in. Note, that because the dynamical equilibrium is
given (approximately) by $\nabla_i {\cal U}_{\rm eff} = \nabla_i
P/\rho$, with $\rho$ being the density, one may also say that it
is the pressure gradient (the pressure stress) that holds the
matter inside $r_{\rm pot}$.

The specific angular momentum in the Novikov-Thorne model is {\it
assumed} to be Keplerian. Slim disk models do not a priori assume
an angular momentum distribution, but self-consistently calculate
it from the relevant equations of hydrodynamics
(\ref{appendix-mass})-(\ref{appendix-energy}). These calculations
reveal that the type of angular momentum distribution depends on
whether accretion rate and viscosity locate the flow in the
disk-like, or the Bondi-like type.

In the {\it Bondi-type} accretion flows the angular momentum is
everywhere sub-Keplerian, ${\cal L} < {\cal L}_K$. These flows are
typical for high viscosities and high accretion rates, as in the
case of $\alpha = 0.1$ and $\dot m = 10$ shown in
Figure~\ref{figure-angular-momentum}. This is the only Bondi-like
flow in this Figure.
%
%
In the {\it disk-like} accretion flows, the angular momentum of
the matter in the disk is sub-Keplerian everywhere, except the
strong-gravity region $r_{\rm pot} < r < r_{\rm cen}$ {\it where
the flow is super-Keplerian}, ${\cal L} > {\cal L}_K$. The radius
$r_{\rm cen}
> r_{\rm ISCO}$ corresponds to the ring of the maximal pressure in
the accretion disk. This is also the minimum of the effective
potential. The radius $r_{\rm pot} < r_{\rm ISCO}$ marks a saddle
point for pressure and effective potential; this is also the
location of the ``potential spout inner edge'', $r_{\rm in} =
r_{\rm pot}$.

Note that in the classic solutions for spherically accretion flows
found by \cite{bondi-1952} the viscosity is unimportant and the
sonic point is saddle, while in the ``Bondi-like'' flows discussed
here, angular momentum transport by viscosity is essentially
important and the sonic point is usually nodal. Therefore, one
should keep in mind that the difference between these types of
accretion flows is also due to the relative importance of pressure
and viscosity. For this reason a different terminology is often
used. Instead of ``disk-like'' one uses the term
``pressure-driven'' and instead  of ``Bondi-like'' one uses
``viscosity-driven'' \citep[see e.g.][]{mat-1984, kat-2008}.

\begin{figure}[h]
\centering
\includegraphics[width=.5\textwidth]{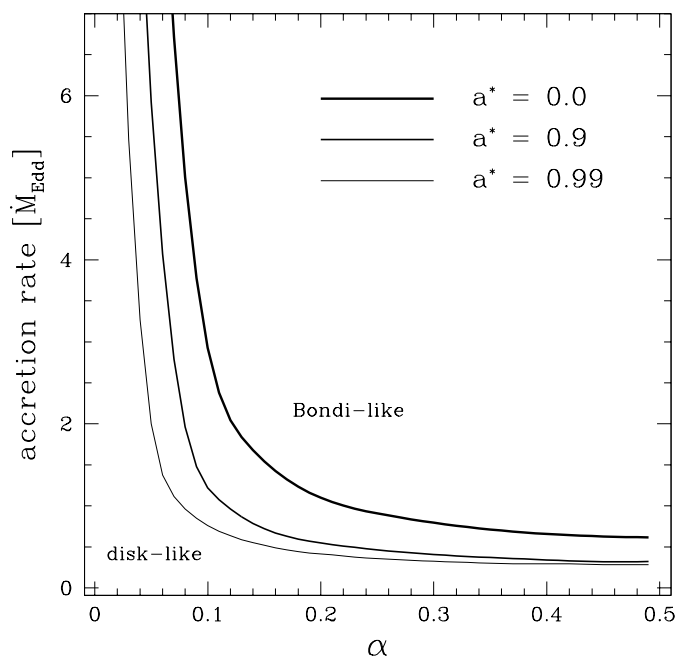}
\caption{Location of the Bondi-like and the disk-like slim
accretion disks in the $[\alpha, {\dot m}]$ parameter space. The
Bondi-like accretion flows are everywhere sub-Keplerian. Disk-like
flows are sub-Keplerian for most radii, but have also a
super-Keplerian part around ISCO.}
\label{figure-disk-bondi}
\end{figure}

From the above discussion it is clear that the location of this
particular inner edge $r_{\rm pot}$ is formally given as the
smaller of the two roots, $r_{\pm} = (r_+, r_-)$, of the equation
\begin{equation}
\label{definition-potential-edge}
\bigl[{\cal L}(r) - {\cal L}_K(r)\bigr]_{r = r_{\pm}} = 0.
\end{equation}
The larger root corresponds to $r_{\rm cent}$. Obviously, equation
(\ref{definition-potential-edge}) has always a solution for the
disk-like flows, and never for the Bondi-like flows. Figure
\ref{figure-disk-bondi} shows a division of the parameter space
into regions occupied by Bondi-like and disk-like flows.
%
%

\begin{figure}[h]
\centering
\includegraphics[width=.5\textwidth,
angle=0]{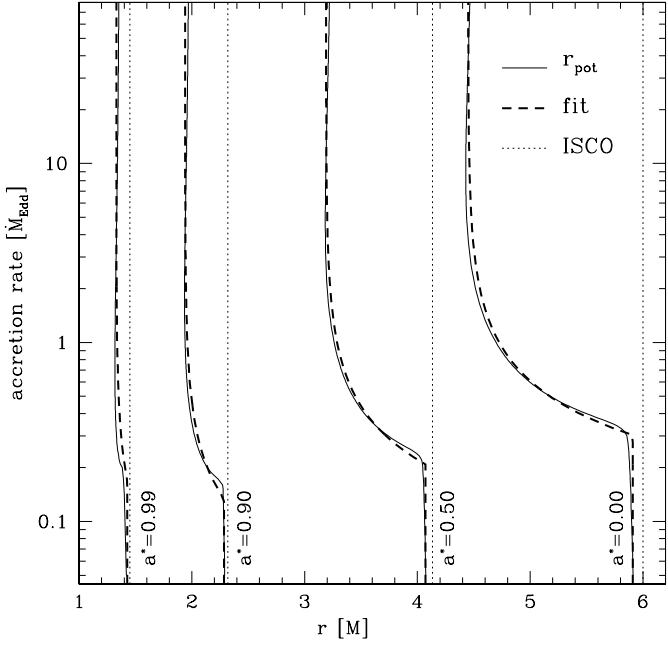}
\caption{Location of the potential spout inner edge $r_{\rm pot}$
for viscosity $\alpha=0.01$ and $a^*=0$. Solid lines show the exact location of
$r_{\rm pot}$ given by equation~(\ref{definition-potential-edge}).
The approximation (\ref{cusp-fitting-formula}) is shown by dashed
lines, and the location of ISCO by dotted lines.}
\label{figure-potential-edge}
\end{figure}

The location of the potential spout inner edge $r_{\rm pot}$ is
shown in Figure~\ref{figure-potential-edge} for $\alpha=0.01$.
Note that for small accretion rates, ${\dot m} \lesssim 0.3$,
location of the potential spout inner edge coincides with ISCO. At
${\dot m} \approx 0.3$, the location of the potential spout jumps
to a new position, which is close to the radius of the innermost
bound circular orbit, $r_{\rm IBCO}$. This behavior is now
well-known. It was noticed first by \cite{koz-1978} for Polish
doughnuts, and by \cite{abr-1988} for slim disks.
%
%
We conclude the Section on the potential spout inner edge by
giving an approximate formula for its location,
\begin{eqnarray}
   r_{\rm pot}(a^*,\dot m) =
   &{\rm Min}\,\bigl[&(0.275 - 0.410a^* + 0.143a^{*2})\dot m^{-1.4}+
   \nonumber \\
  &&4.45 - 4.87a^* + 8.06a^{*2} - 6.38a^{*3}\,;\nonumber \\
  &&0.985\,r_{\rm ISCO}\,\,\bigr].
  \label{cusp-fitting-formula}
\end{eqnarray}
The formula (\ref{cusp-fitting-formula}) is valid for $\alpha =
0.01$.
%
%
%
\section{The sonic edge}
\label{section-sonic}
%

By a series of algebraic manipulations one reduces the slim disk
equations (\ref{appendix-mass})-(\ref{appendix-energy}) to a set
of two ordinary differential equations for two dependent
variables, e.g. the Mach number $\eta=-V/c_S^2=-V\Sigma/P$ and the
angular momentum ${\cal L}=-u_\phi$,
\be
\der{\ln \eta}{\ln r}=\frac{{\cal N}_1(r,\eta,{\cal L})}{{\cal D}(r,\eta,{\cal L})}\\
\der{\ln L}{\ln r}=   \frac{{\cal N}_2(r,\eta,{\cal L})}{{\cal
D}(r,\eta,{\cal L})}
\label{eq_reg}
\ee
%
%
\begin{figure}[h]
\centering
\includegraphics[height=.5\textwidth, angle=-90]{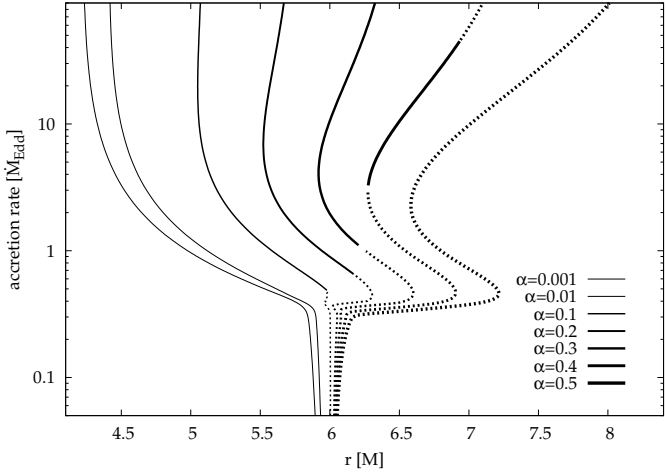}
\caption{Location of the sonic point as a function of the
accretion rate for different values of $\alpha$, for a
non-rotating black hole, $a^* = 0$. The solid curves are for saddle
type solutions while the dotted curves present nodal type
regimes.}
\label{fig:sonic-point-location}
\end{figure}
For a non-singular physical solution the nominators ${\cal N}_1$
and ${\cal N}_2$ must vanish at the same radius as the denominator
${\cal D}$. The denominator vanishes at the sonic edge (or sonic
radius) where the Mach number is close to unity, i.e.
\begin{equation}
{\cal D}(r,\eta, {\cal L})_{\vert \,r = r_{\rm son}} = 0.
\label{definition-sonic}
\end{equation}
For low mass accretion rates, smaller than about $0.3{\dot
M_{\rm Edd}}$ in case of $a^*=0$, the sonic edge $r_{\rm son}$ locates
close to ISCO, independently on the viscosity $\alpha$, as
Figure~\ref{fig:sonic-point-location} shows. At about $0.3{\dot
M_{\rm Edd}}$ a qualitative change occurs, resembling a ``phase
transition'' from the Shakura-Sunyaev behavior, to a very
different slim-disk behavior.

For higher accretion rates the location of the sonic point
significantly departs from ISCO. For low values of $\alpha$ the
sonic point moves closer to the horizon down to $\sim4M$ for
$\alpha=0.001$. For $\alpha>0.2$ the sonic point moves outward
with increasing accretion rate reaching values as high as $8M$ for
$\alpha=0.5$ and $100\dot M_{\rm Edd}$. This effect was first
noticed for small accretion rates by \cite{muchotrzeb-1986} and
later investigated in a wide range of accretion rates by
\cite{abr-1988}, who explained it in terms of the disk-Bondi
dichotomy. The dependence of the sonic point location on the
accretion rate in the near-Eddington regime is more complicated
and is related to the fact that in this range of accretion rates
the transition from the radiatively efficient disk
to the slim disk occurs near the sonic radius.

The topology of the sonic point is important, because physically
acceptable solutions must be of the saddle or nodal type; the
spiral type is forbidden. The topology may be classified by the
eigenvalues $\lambda_1, \lambda_2, \lambda_3$ of the Jacobi
matrix,
\be
{\cal J} = \left[
\begin{array}{ccc}
 \pder{\cal D}r      &\pder{\cal D}\eta      &\pder{\cal D}{\cal L}\\
 \pder{{\cal N}_1}r  &\pder{{\cal N}_1}\eta  &\pder{{\cal N}_1}{\cal L}\\
 \pder{{\cal N}_2}r  &\pder{{\cal N}_2}\eta  &\pder{{\cal N}_2}{\cal L}
\end{array}
\right].
\ee
Because ${\rm det}({\cal J}) =0$, only two eigenvalues $\lambda_1,
\lambda_2$ are non-zero, and the quadratic characteristic equation
that determines them takes the form,
\be
2\,\lambda^2 - 2\,\lambda\,{\rm tr}({\cal J}) - \left[{\rm
tr}({\cal J}^2) - {\rm tr}^2({\cal J})\right] = 0.
\ee
The nodal type is given by $\lambda_1\lambda_2 > 0$ and the saddle
type by $\lambda_1\lambda_2 < 0$, as marked in
Figure~\ref{fig:sonic-point-location} with the dotted and the
solid lines, respectively. For the lowest values of $\alpha$ only
the saddle type solutions exist. For moderate values of $\alpha$
($0.1\le\alpha\le0.4$) the topological type of the sonic point
changes at least once with increasing accretion rate. For the
highest $\alpha$ solutions have only nodal type critical points.

\begin{figure}[h]
\centering
\includegraphics[height=.5\textwidth,
angle=-90]{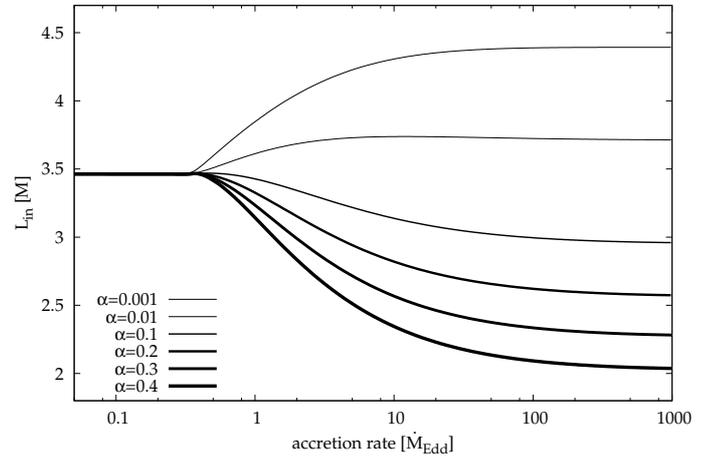}
\caption{Angular momentum at the horizon dependence on accretion
rate for solutions with different values of $\alpha$ for $a^*=0$.}
\label{fig:horizon-angular-momentum}
\end{figure}

The extra regularity conditions at the sonic point ${\cal
N}_i(r,\eta,{\cal L}) = 0$ are satisfied only for one particular value of
the angular momentum at the horizon  which is the {\it eigenvalue}
of the problem. ${\cal L}_{in}$ is not known a priori, and should be
found. Figure~\ref{fig:horizon-angular-momentum} shows how does
${\cal L}_{in}$ depend on the accretion rate and the $\alpha$
viscosity parameter.
%
%

%
\section{The variability edge}
\label{section-QPO-variability}
%

Axially symmetric and stationary states of slim accretion disks
represent, obviously, only a certain theoretical idealization.
Real disks are non-axial and non-steady. In particular, one
expects transient coherent features at accretion disk surfaces
--- clumps, flares, and vortices. Orbital motion of these features
could quasi-periodically modulate the observed flux of radiation,
mostly through the Doppler effect and the relativistic beaming.
Let $\Pi$ be the ``averaged'' variability period, and $\Delta \Pi$
a change of the period during one period due to radial motion of a
spot.
%
%
\begin{figure*}[t]
\centering
\includegraphics[height=7.0cm]{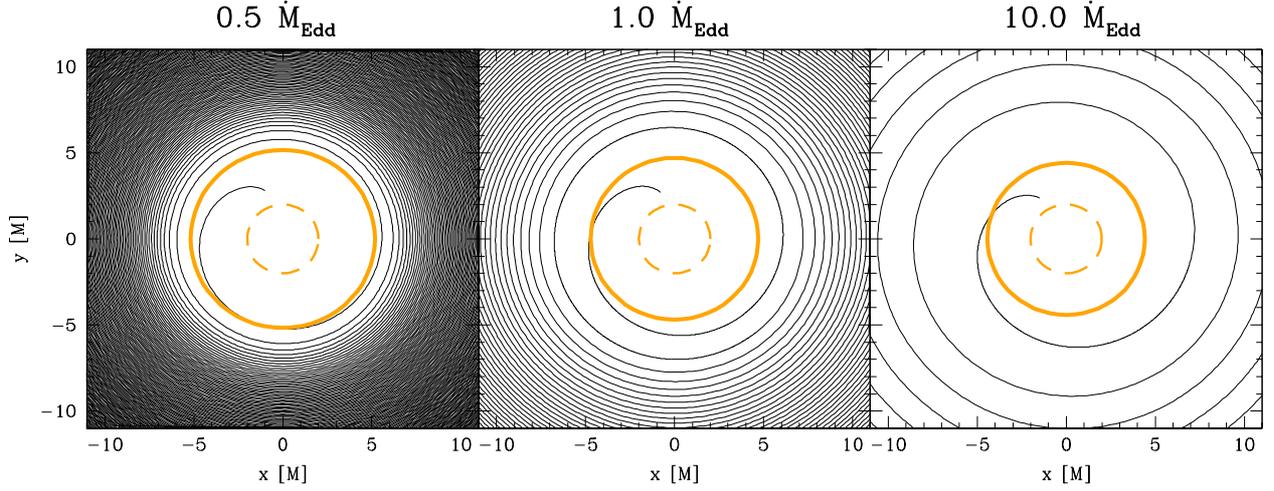}
\caption{The fluid
flow trajectories in slim accretion disks shown by thin solid
lines for different accretion rates. Locations of $r_{\rm pot}$
and the location of
black hole horizon are shown by thick gray solid and broken lines, respectively.
For small accretion
rates, the pattern of trajectories consists of very tight spirals
(almost circles) for $r > r_{\rm pot}\approx r_{\rm ISCO}$ and very wide spirals
(almost a radial fall) for  $r < r_{\rm pot}$. In this case,
there is a sharp transition from almost circular motion to almost
radial free-fall that clearly defines the variability edge as
$r_{\rm var} = r_{\rm ISCO}.$ For higher accretion rates, the
fluid trajectories are wide open spirals in the whole inner region
of the flow and the variability edge makes no sense.}
\label{figure-spirals}
\end{figure*}
The variability quality factor $Q$ may be estimated by,
\begin{equation}
\frac1Q=\frac{\Delta
\Pi}{\Pi}=
\frac{\Delta\Omega}{\Omega}=\frac1{\Omega}\der\Omega{r}\Delta
r
= 2\pi\frac1{\Omega^2}\der\Omega{r}\frac{u^r}{u^t}
\label{quality-factor-definition}
\end{equation}
where $u^r/u^t = dr/dt$ and $u^r$ and $u^t$ are contravariant
components of the four velocity. The period relates to the orbital
angular velocity by $\Pi = 2\pi/\Omega$. Using the relations (see
Appendix A for the explanation of the notation used),
\begin{eqnarray}
u^r &=& \frac V{\sqrt{1-V^2}}\frac{\sqrt{\Delta}}r\nonumber \\
u^t &=& \frac{\gamma\sqrt A}{r\sqrt\Delta}=\frac{\sqrt
A}{r\sqrt\Delta}\frac1{\sqrt{(1-V^2)(1-({\tilde V}^\phi)^2)}}
\label{contravariant-velocities}
\end{eqnarray}
with $V$ being the radial velocity as measured by an observer
corotating with the fluid, one obtains:
\begin{equation}
Q = \frac{1}{2\pi}\left\vert\der{\log\Omega}{\log
r}\right\vert^{-1} \left\vert\frac{{\bar
V}^\phi}{V}\right\vert\,f^*(a^*, r),
\label{quality-final}
\end{equation}
where,
\begin{eqnarray}
f^*(a^*, r) &\equiv& \frac{r^3}{\sqrt{\Delta A}}
= +\left[ 1 - X - X^2a^{*2}\,(a^{*2} + 1) - X^5\,a^{*4}
\right]^{-1/2}, \nonumber \\
{\bar V}^{\phi} &=& \frac{V^{\phi}}{\sqrt{1-(V^\phi-\omega\tilde
R)^2}},
\label{quality-two-functions}
\end{eqnarray}
with $X=2r_G/r$. From (\ref{appendix-Delta}) and
(\ref{appendix-A}) it is clear than $\Delta A > 0$ outside the
black hole horizon. Note that in Newtonian limit it is $X \ll 1$
and one has $f^*(a^*, r)=1$. In this limit $V$, ${\bar V}^{\phi}$
are the radial and azimuthal component of velocity, and the
formula (\ref{quality-final}) takes its obvious Newtonian form.
%
%
%
\begin{figure}[h]
\centering
\includegraphics[width=.5\textwidth,
angle=0]{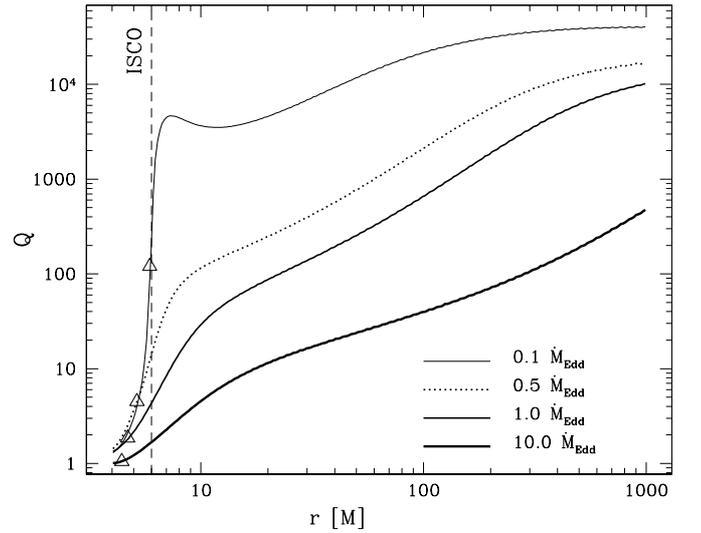}
\caption{The quality factor $Q$ profiles for different accretion
rates. Triangles show $r_{\rm pot}$ for each rate.
The vertical dashed line denotes the location of ISCO.}
\label{figure-quality-factor-Q}
\end{figure}

Behavior of the quality factor $Q$ is shown in Figure
\ref{figure-quality-factor-Q}. Profiles for four accretion rates
are drawn. As Fig.~\ref{figure-spirals} shows the lower accretion
rate the smaller radial velocity component and therefore the
quality factor $Q$ in general increases with decreasing accretion
rate. For the lowest values of $\dot m$ a rapid drop is visible
at ISCO corresponding to the change in the nature of the flow
(gas enters the free-fall region below ISCO). For higher accretion
rates such behaviour is suppressed as the trajectories become
wide open spirals well outside ISCO.

Note that our definition (\ref{quality-factor-definition}) of the
quality factor $Q$, essentially agrees with a practical definition
of the variability quality factor $Q_0$ defined by observers with
the help of the observationally constructed Fourier variability
power spectra, $I(\nu)$. Here $I(\nu)$ is the observed variability
power (i.e. the square of the observed amplitude) at a particular
observed variability frequency $\nu$. Any observed {\it quasi}
periodic variability with the frequency $\sim \nu_0$ shows in the
power spectrum as a local peak in $I(\nu)$, centered at a certain
frequency $\nu_0$. The half-width $\Delta \nu$ of the peak defines
the variability quality factor by $Q_0 = \nu/\Delta \nu_0$.

Quasi periodic variability with kHz frequencies, called kHz QPO,
is observed from several low-mass neutron star and black hole
binaries. In a pioneering and important research, \cite{barr-2005}
carefully measured the quality factor for a particular source in
this class (4U 1608-52) and found that $Q_0 \sim 200$, i.e. that
the kHz are very coherent. They argued that $Q_0 \sim 200$ cannot
be due to kinematic effects in orbital motion of hot spots, clumps
or other similar features located at the accretion disk surface,
because these features are too quickly sheared out by the
differential rotation of the disk \citep[see also][]{bath-1974,
pringle-1981}. They also argued that although coherent vortices
may survive much longer times at the disk surface
\citep[e.g.][]{abr-nature-1992}, if they participate in the inward
radial motion, the observed variability $Q_0$ cannot be high. Our
results shown in Figure~\ref{figure-quality-factor-Q} illustrate
and strengthen this point. We also agree with the conclusion
reached by \cite{barr-2005} that the observational evidence
against orbiting clumps as a possible explanation of the
phenomenon of kHz QPO, seems to point out that this phenomenon is
most probably due to the accretion disk global
oscillations\footnote{\cite{barr-2005} found also how $Q_0$ varies
in time for each of the two individual oscillations in the
``twin-peak QPO''. This gives strong observational constraints for
possible oscillatory models of the twin peak kHZ QPO; see also
\cite{boutelier-2010}.}. For excellent reviews of the QPO
oscillatory models see \cite{wagoner-1999} and \cite{kato-2001}.

Although clumps, hot-spots, vortices or magnetic flares cannot
explain the coherent kHz QPOs with $Q_0 \sim 200$, they certainly
are important in explaining the continuous, featureless Fourier
variability power spectra \citep[see e.g.][and references quoted
there]{abr-bao-1991, schnittman05, pech-2008}. Our results shown in
Figure~\ref{figure-quality-factor-Q} indicate that: (i) at low
accretion rates, a sharp high-frequency cut-off in $I(\nu)$ may be
expected at about the ISCO frequency, (ii) at high accretion rates
there should be no cut-off in $I(\nu)$ at any frequency, (iii) the
logarithmic slope $p = (d\ln I/d\ln \nu)$ should depend on ${\dot
m}$.

A more quantitative description of (i)-(iii) will be given in a
future publication \citep{str-variability-2010}.

%
\section{The stress edge}
\label{section-stress}
%

The Shakura-Sunyaev model {\it assumes} that there is no torque at
the inner edge of the disk, which in this model coincides with
ISCO. Slim disk model {\it assumes} that there is no torque at the
horizon of the black hole. It makes no assumption on the torque at
the disk inner edge, but calculations {\it prove} that the torque
is small there.

\begin{figure}
\centering
 \subfigure
{
\includegraphics[height=.5\textwidth, angle=270]{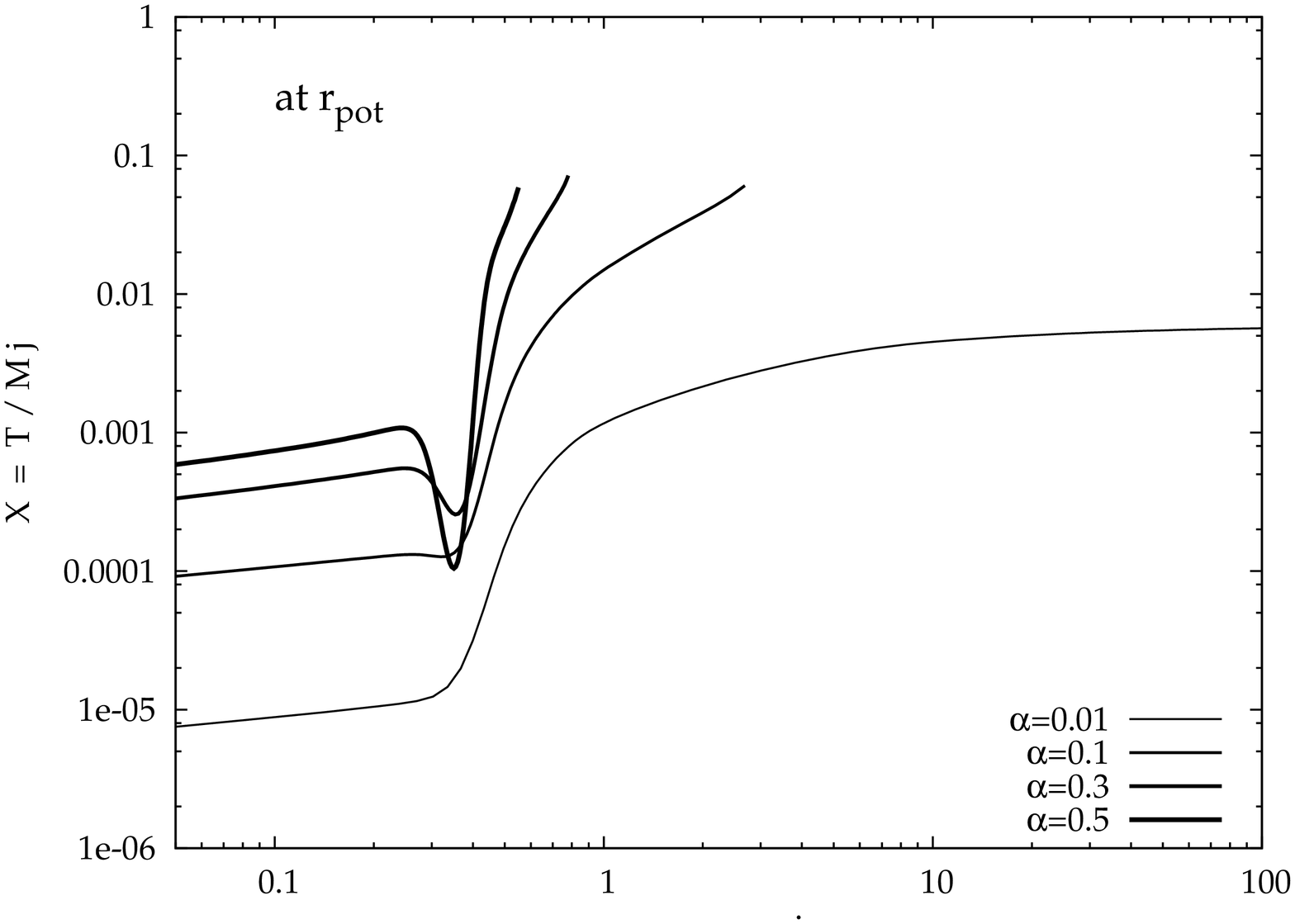}
}
\\
\vspace{-.02\textwidth}
 \subfigure
{
\includegraphics[height=.5\textwidth, angle=270]{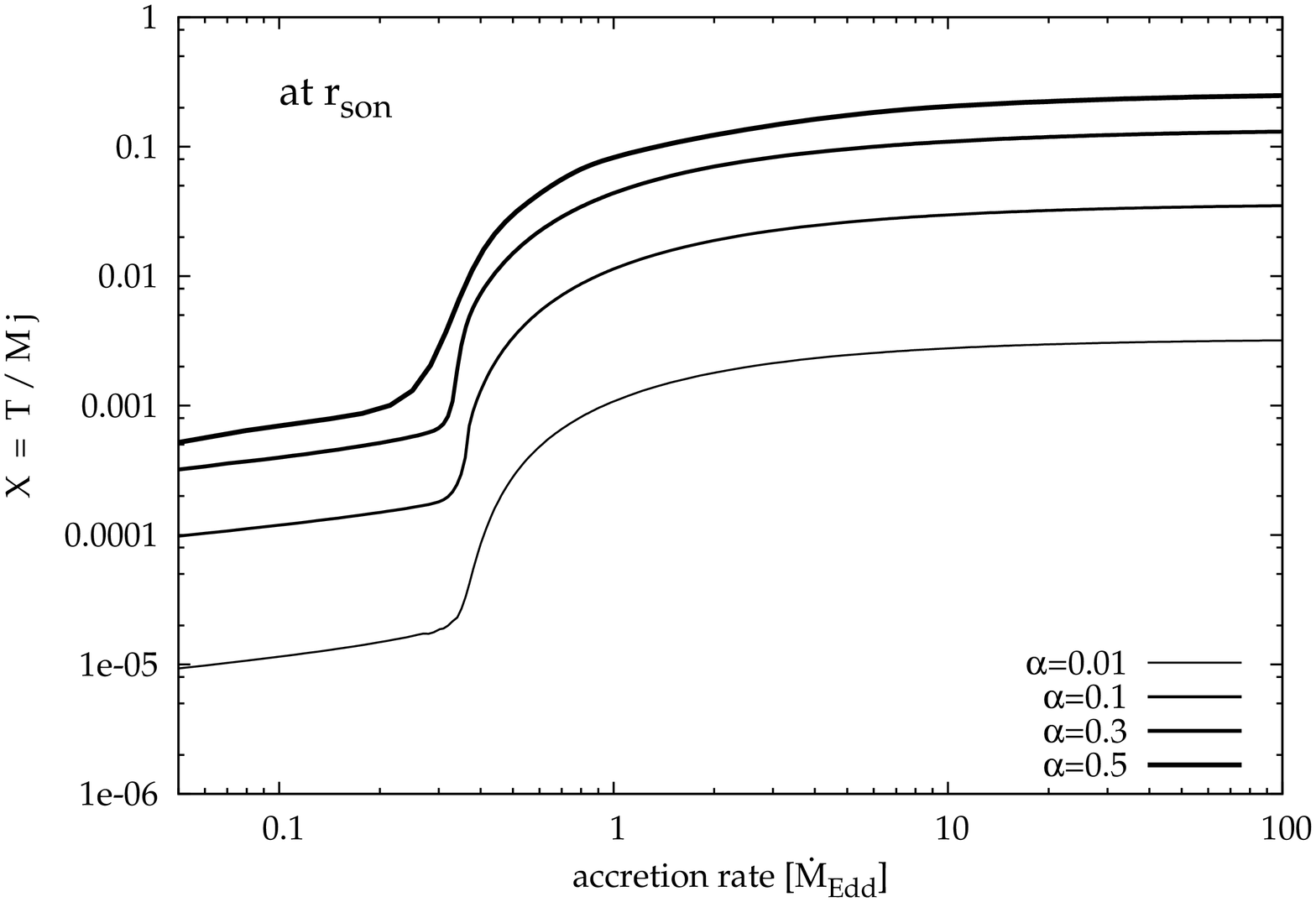}
}
\caption{Ratio of the angular momentum flux caused by torque to
the flux caused by advection calculated at $r_{\rm pot}$ (top) and
$r_{\rm son}$ (bottom panel) versus mass accretion
rate for a number of values of $\alpha$ and $a^*=0$. The $r_{\rm pot}$
profiles for high viscosities terminate when disk enters the
Bondi-like regime.}
\label{fig:inner-torque}
\end{figure}

The zero-torque at the horizon is consistent with the small torque
at the inner edge of slim disks, as Figure \ref{fig:inner-torque}
shows.
%
%
The Figure presents the relative importance of the torque ${\cal
T}$ by comparing it with the ``advective'' flux of angular
momentum ${\dot M}j$ (c.f. equation \ref{torque-definition}). For
the viscosity parameter $\alpha$ smaller than about $0.01$, the
ratio $\chi = {\cal T}/{\dot M}j$ both at $r_{\rm pot}$ and $r_{\rm son}$
is smaller than $0.01$ even for highly super-Eddington accretion
rates, and for small accretion rates the ratio is vanishingly
small, $\chi \approx 10^{-5}$. For high viscosity, $\alpha = 0.5$,
the ratio is very small for small accretion rates, $\chi <
10^{-3}$ and still smaller than about $0.1$ even for
super-Eddington accretion rates (calculated at the sonic radius as
the disk enters the Bondi-like regime for such high accretion
rates).
%
%
\begin{figure}[h]
\centering
\includegraphics[width=.5\textwidth, angle=0]{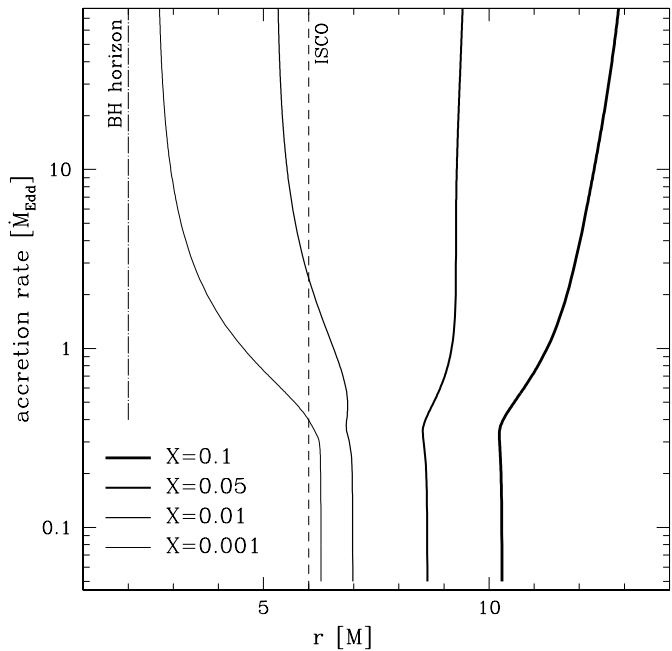}
\caption{Profiles of $r_{\rm str}$ defined as the radius with given
value of the torque parameter $\chi$ for $\alpha=0.01$. BH horizon and ISCO are also
shown with dot-dashed and dashed lines, respectively.}
\label{fig:rstr}
\end{figure}

To define the \textit{stress inner edge} $r_{\rm str}$ one has to
specify the characteristic value of the torque parameter ${\chi}$.
Profiles of $r_{\rm str}$ for a few values of ${\chi}$ and $\alpha=0.01$ are shown in
Fig.~\ref{fig:rstr}. The stress edge for ${\chi} \rightarrow 0$ is
located at ISCO for low accretion rates. When accretion rate
exceeds $\sim 0.3\dot M_{\rm Edd}$ it departs from ISCO and moves
closer to BH approaching its horizon with increasing $\dot m$.
Behaviour of $r_{\rm str}$ profiles for higher ($\gtrsim 0.1$) values
of ${\chi}$ is different - they move away from the BH as the
angular momentum profiles become flatter with increasing accretion
rates (compare Fig.~\ref{figure-angular-momentum}).

In the case of disk-like accretion with a {\it very low viscosity}
$\alpha \le 0.01$, it is with high accuracy,
\be
\label{definition-unique}
r_{\rm pot} \approx r_{\rm son}.
\ee
In this case the ``inner edge'' inherits both the sonic edge and
the potential spout edge properties; suggesting a small torque. It
looks, as this is indeed the case. By pushing the MHD numerical
simulations to their limits, \cite{sha-2008} and recently \cite{Pennaetal-10}
 calculated a thin,
$H/r \lesssim 0.1$, disk-like accretion flow, and showed that for it
the inner edge torque was small.
%
\section{The radiation edge}
\label{section-radiation}
\begin{figure}[h]
\centering
\includegraphics[width=.5\textwidth, angle=0]{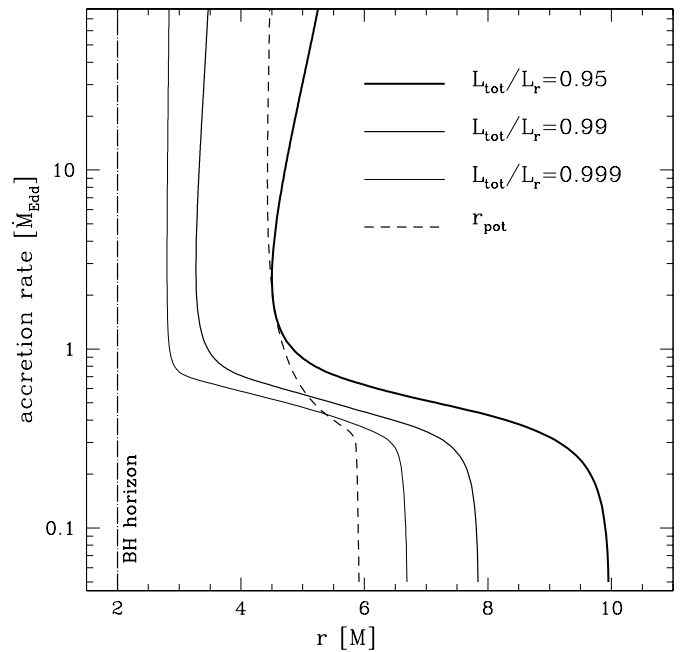}
\caption{``Luminosity edges'' defining inner radii of area emitting
given amount of the total disk radiation. The lines are drawn
for 95\%, 99\% and 99.9\% of the total emission. The dashed line
shows the location of the \textit{potential spout} inner edge $r_{\rm pot}$.
The gravitational suppression of the radiation has been taken into
account.}
\label{fig:radiation-radius}
\end{figure}

%
As discussed in the previous section, the torque at $r_{pot} <
r_{\rm ISCO}$ is small, but non-zero and therefore there is
orbital energy dissipation also at radii smaller than ISCO. Thus,
some radiation from this region takes place and the inner edge is
not expected to coincide with the radiation edge, $r_{\rm rad} <
r_{\rm pot}$. In Fig.~\ref{fig:radiation-radius} we present profiles
of $r_{\rm rad}$ defined as radii limiting area emitting given
fraction of disk total luminosity. For low accretion rates
($<0.3\dot M_{\rm Edd}$) disk emission terminates close to ISCO as the
classical models of accretion disks predict. Locations of the
presented $r_{\rm rad}$ are determined by the regular Novikov \&
Thorne flux radial profile. For higher accretion rates disk
becomes advective and the maximum of the emission moves
significantly inward. As a consequence of the increasing rate of
advection (and resulting inward shift of $r_{\rm pot}$) the efficiency
of accretion drops down.

We want to stress here that the location of the radiation edge is
{\it not} determined by the location of the stress edge (as some
authors seem to believe), but by the fact that significant
advection flux brings energy into the region well below ISCO.

Let $r_{\rm out} \gg r_{\rm G}$ be the outer radius of the disk.
The total luminosity of the disk could be estimated from
\begin{eqnarray}
L &=& {\dot M}e_{\rm rad} + (Q\,\Omega)_{\rm rad}
- {\dot M}e_{\rm out} - (Q\,\Omega)_{\rm out} \nonumber \\
0 &=& {\dot M}{\cal L}_{\rm rad} + Q_{\rm rad}
- {\dot M}{\cal L}_{\rm out} - Q_{\rm out}.
\label{luminosity-estimate}
\end{eqnarray}
It is $\Omega_{\rm out} \approx 0$, $e_{\rm out} \approx 0$, and
from this one derives
\begin{equation}
L = {\dot M}\left[ e_{\rm rad} + \chi ( {\cal L}\Omega )_{\rm rad}
\right] \equiv \eta {\dot M}
\label{efficiency-estimate}
\end{equation}
where $\chi$ is the ratio of the viscous torque to the advective
flux of angular momentum (see Figures \ref{fig:inner-torque} and
\ref{fig:rstr}).

Because $\chi \ll 1$, the efficiency of accretion $\eta$ depends
mainly on the specific energy at the inner edge, $e_{\rm rad}$.
The further away is the inner edge from ISCO (and closer to the
black hole), the smaller is the efficiency.
%
%

%
\section{The reflection edge}
\label{section-reflection}
%

The iron K$_{\alpha}$ fluorescent line is one of the
characteristic features observed in many sources with black hole
accretion disks \citep{Miller-2006,Remillard-2006}. The intensity and the shape
of this line depends strongly on the physical conditions close to
the inner edge.
This was discussed by many authors, including
\cite{rey-2008} who gave three conditions for line formation: (i)
the flow has to be Thomson-thick in the vertical direction; (ii)
disk has to be irradiated by external source of X-rays (hard X-ray
irradiation plays
crucial role in the process of fluorescence and changes the
ionization degree of matter); (iii) the ionization
state should
 be sufficiently low (iron cannot be fully ionized).

\begin{figure}
\centering
 \subfigure
{
\includegraphics[height=.5\textwidth, angle=270]{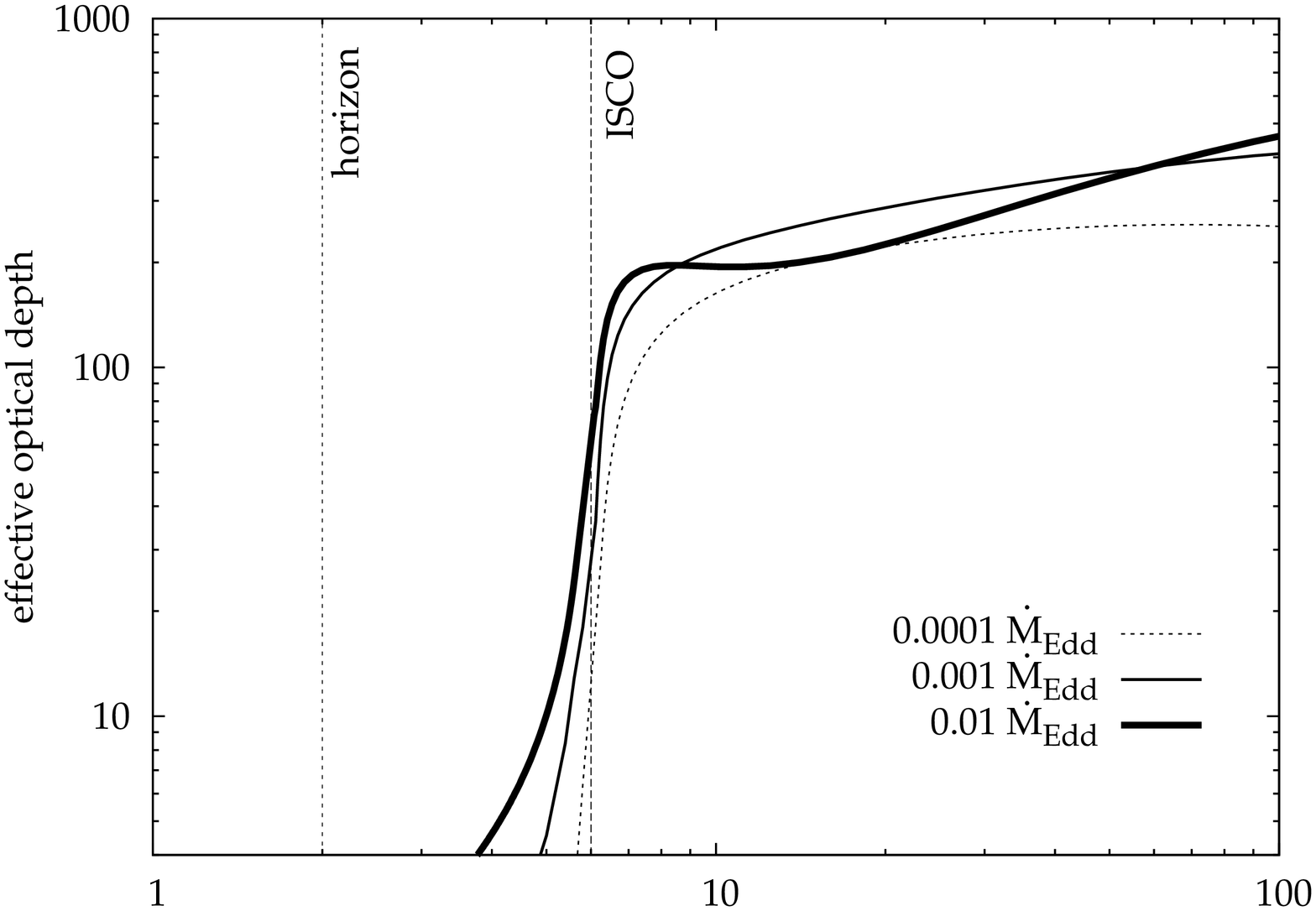}
}
\\
\vspace{-.02\textwidth}
 \subfigure
{
\includegraphics[height=.5\textwidth, angle=270]{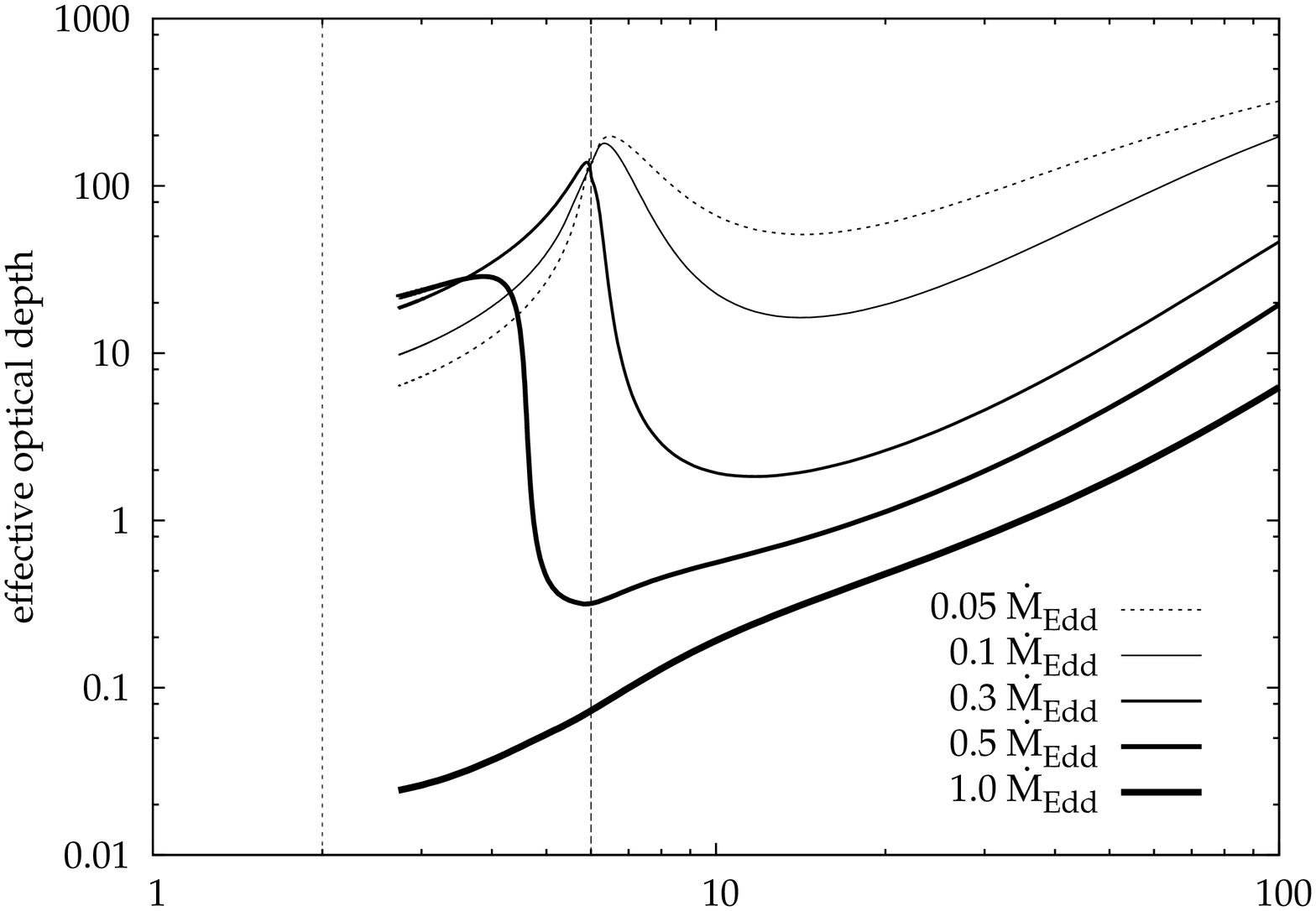}
}
\\
\vspace{-.02\textwidth}
 \subfigure
{
\includegraphics[height=.5\textwidth, angle=270]{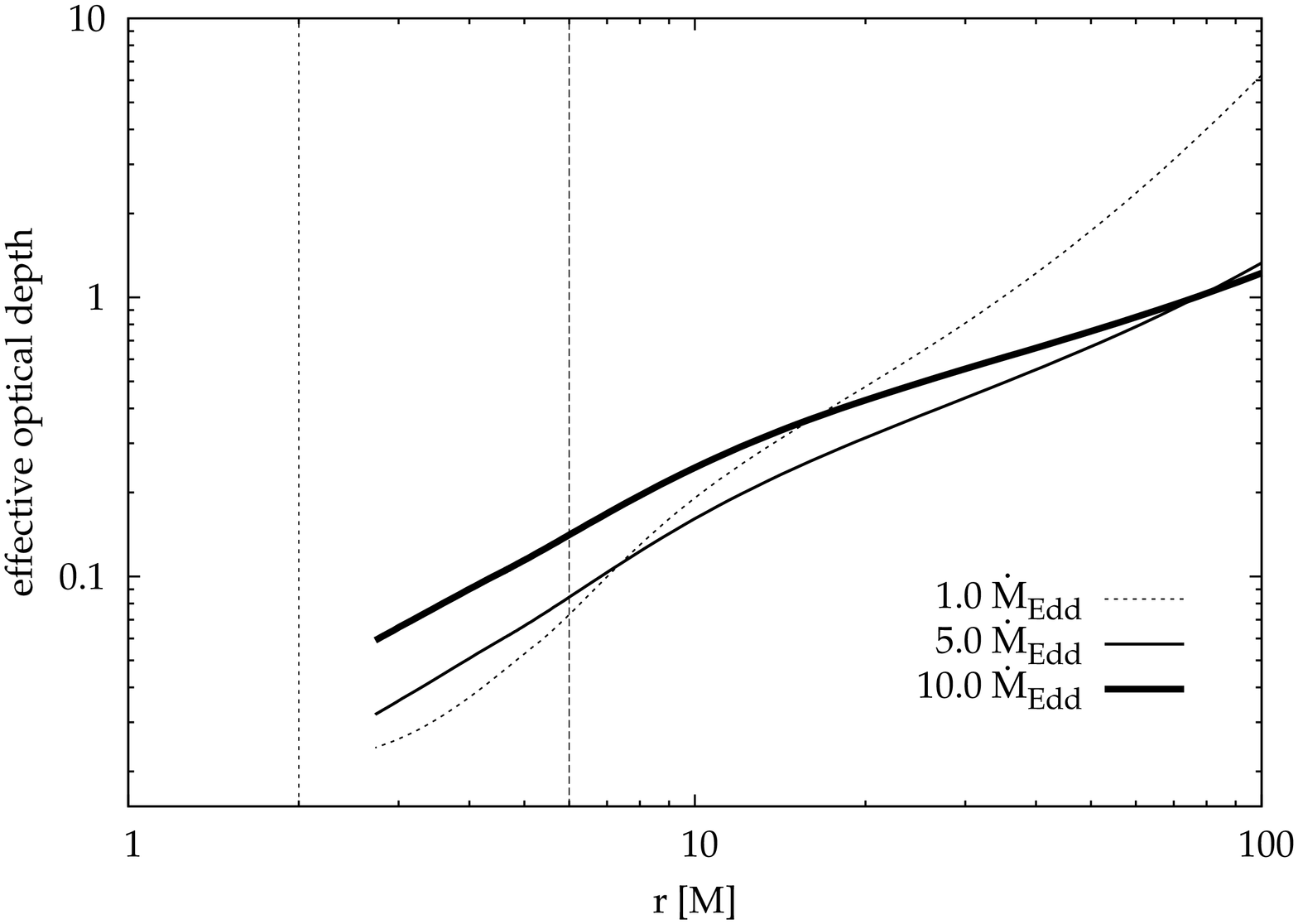}
} \caption {Profiles of the effective optical depth $\tau_{\rm eff}$
for $\alpha=0.1$ and $a = 0$ in three different regimes of
accretion rate. Vertical lines denote locations of the BH horizon
(dotted) and ISCO (dashed). Three types of behavior of
$\tau_{\rm eff}$ are seen: {\it sharp drop} at ISCO for smallest
accretion rates, {\it maximum} near ISCO for moderate accretion
rates, and {\it monotonic} everywhere for highest accretion rates.
}
\label{fig:tau_a0}
\end{figure}

We point out here, that this first condition is sufficient for the
formation of the reflection continuum, but formation of the
fluorescent iron line requires an even stronger condition i.e.
that the effective optical depth of the flow should be higher than unity:

\begin{equation}
\label{effective-depth}
\tau_{\rm eff} = \sqrt{\tau_{\rm abs} (\tau_{\rm abs}+\tau_{\rm es})} > 1.
\end{equation}
This is because, the fluorescence requires efficient absorption of
high energy photons by iron ions. In Fig.~\ref{fig:tau_a0} we
present profiles of the effective optical depth $\tau_{\rm eff}$
in different regimes of accretion rates for $\alpha=0.1$ and $a^*
= 0$. Three characteristic types of their behaviour are shown:
\textit{sharp drop}, \textit{maximum} and \textit{monotonic} at
the top, middle and bottom panels, respecively. Behaviour for
different values of $\alpha$ and $a^*$ is qualitatively similar
(but not quantitatively as in general $\tau_{\rm eff}$ increases
with decreasing $\alpha$). Top panel, corresponding to the lowest
accretion rates, shows a {\it sharp drop in} $\tau_{\rm eff}$ near
ISCO. The same behavior was noticed previously e.g. by
\cite{rey-2008}. The drop could clearly define the inner
reflection edge $r_{\rm ref}\approx r_{ISCO}$ limiting the radii
where formation of the fluorescent iron line is prominent. The
middle panel, corresponding to moderate accretion rates, shows a
{\it maximum in} $\tau_{\rm eff}$ near ISCO. The non-monotonic
behaviour is caused by the fact that regions of moderate radii
outside ISCO become radiation pressure and scattering dominated.
Note, that the top of the maximum of $\tau_{\rm eff}$ stays near
ISCO in a range of accretion rates, but for accretion rates
greater than $0.3\dot M_{\rm Edd}$ it moves closer to the black
hole with increasing ${\dot m}$ as the disk emission profile
changes due to advection. The bottommost panel corresponds to
super-Eddington accretion rates. The profiles are {\it monotonic
in} $\tau_{\rm eff}$ and define no characteristic inner reflection
edge. Close to the black hole such flows are effectively optically thin
reaching $\tau_{\rm eff}=1 $ on about few tens of gravitational
radii.

When effective optical depth of the flow becomes less then unity,
our approximation of radiative transfer by diffusion with grey
opacities (Eq.~\ref{appendix-energy}) becomes not valid. In such
case full radiative transfer through accretion disks atmospheres
should be solved \citep[e.g.][]{davi-2005,agatamadej08}. Still,
our results allow us to estimate roughly how far from the black
hole the iron line formation is most prominent, assuming that disk
is uniformly illuminated by an exterior X-ray source. For accretion rates
smaller than $~0.3\dot M_{\rm Edd}$, the reflection edge is located
very close to ISCO and we may identify shape of the iron line
with gravitational and dynamical effects connected to ISCO. In case of higher but
sub-Eddington accretion rates, the maximum of the effective
optical depth is located inside ISCO what may possibly allow us to study
extreme gravitational effects on the iron line profile. However, the
assumption that the line is formed at ISCO is no longer satisfied.
The super-Eddington flows have smooth and monotonic profiles
of the effective optical depth. Therefore, the reflection edge cannot
be uniquely defined and no relation between shape of the
fluorescent lines and ISCO exists. Finally, one should keep in mind that such lines can be
successfully modeled by clumpy absorbing material and may have nothing to do
with relativistic effects \citep[see e.g.][and references therein]{Milleretal-2009}.
The role of the ISCO in determining the
shape of the Fe lines was also questioned in the past (based on different
reasoning) by \cite{ReynoldsBegelman-97} whose arguments were then refuted by
\cite{Youngetal-98}.

%
%

%
\section{Conclusions}
\label{section-conclusions}
%

We addressed the inner edge issue by discussing behavior of six
differently defined ``inner edges'' of slim accretion disks around
the Kerr black hole. We found that the slim disk inner edges
behave very differently than the corresponding Shakura-Sunyaev and
Novikov-Thorne ones. The differences are qualitative. Even for
moderate luminosities, ${\dot M} \gtrsim 0.3\,{\dot M}_{\rm Edd}$,
there is no unique inner edge. Differently defined edges locate at
different places. For nearly Eddington luminosities, the
differences are huge and the notion of the inner edge losses all
practical significance.

We summarize the properties and locations of the six inner edges
in Table \ref{table:1}. It refers to $a^*=0$, but the qualitative
behavior is similar for $a^*\not =0$.
%
\begin{table*}[t]
\caption{Summary of the results (specific numbers refer to the
case $a^*=0$).}              
\label{table:1}      
\centering                           
\begin{tabular}[width=1.0\textwidth]{l m{3.cm} m{1.7cm} m{1.25cm} m{2.cm} m{2.cm} m{2.5cm} m{.0001cm}}        
\hline\hline                 
  & \centering$r_{\rm pot}$ & \centering$r_{\rm son}$ & \centering$r_{\rm var}$ & \centering$r_{\rm str}$ & \centering$r_{\rm rad}$ & \centering $r_{\rm ref}$ &\\
\hline                        
&&&&&&\\
 $\dot m\lesssim 0.3$
&
 \multicolumn{6}{c}{
$ r_{\rm in} \approx
r_{\rm pot} \approx
r_{\rm son} \approx
r_{\rm var} \approx
r_{\rm str} \approx
r_{\rm rad} \approx
r_{\rm ref} \approx
r_{\rm ISCO}
$
}
\\&&&&&&\\
\hline                        
 $\dot m\gtrsim 0.3$ &
\begin{center}
for $\alpha\lesssim0.1$ moves inward with increasing $\dot m$ down to $\sim r_{mb}$;

~

for $\alpha\gtrsim0.1$ and sufficiently high $\dot m$ disk enters the Bondi regime --- undefined
\end{center}&
\begin{center}
departs from ISCO;

~

for $\alpha\ll0.1$

$r_{\rm son}\approx r_{mb}$;

~

for $\alpha\gtrsim0.2$

$r_{\rm son}>r_{ISCO}$ \end{center}
&
\begin{center}
undefined\end{center}
&
\begin{center}
moves inward with increasing $\dot m$ down to BH horizon.\end{center}
&
\begin{center}
moves inward with increasing $\dot m$ down to BH horizon.\end{center}
&
\begin{center}
for $0.3\lesssim\dot m\lesssim1.0$ $r_{\rm ref}<r_{ISCO}$

~

for $\dot m\gtrsim1.0$

undefined\end{center}
&
\\
\hline                                    
\end{tabular}
\end{table*}

\begin{figure}[h]
\centering
\includegraphics[width=.5\textwidth]{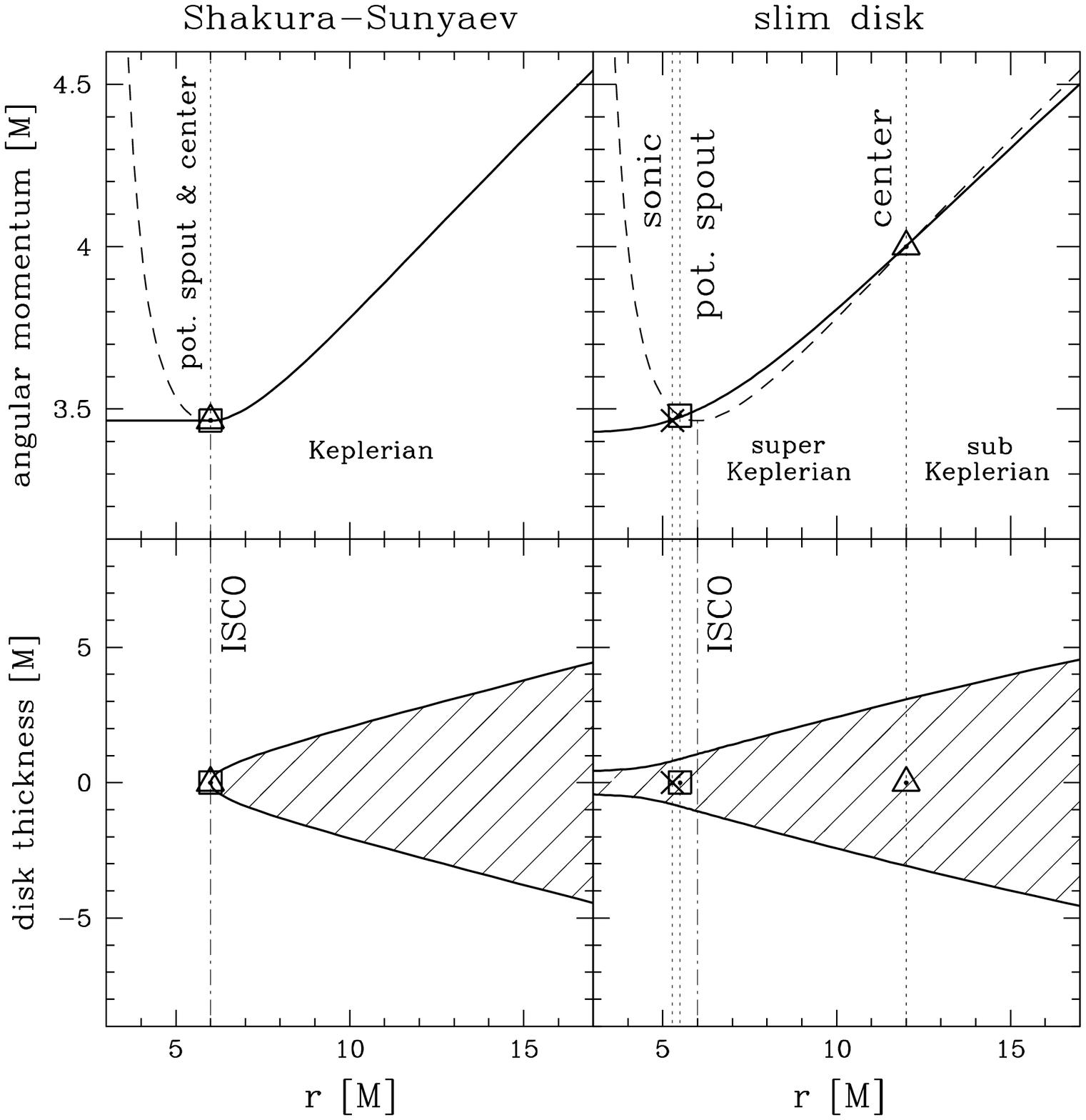}
\caption{The differences between Shakura-Sunyaev and slim disk
picture of the disk inner edge (see text for a detailed
explanation of the Figure).}
\label{fig:inner-ss-versus-slim}
\end{figure}

We conclude, by showing in Figure \ref{fig:inner-ss-versus-slim}
differences between the Shakura-Sunyaev and slim-disk (in the
disk-like case) treatment of the inner disk physics. The innermost
part of a Shakura-Sunyaev disk is shown in the left column in
Figure \ref{fig:inner-ss-versus-slim}, and the innermost part of a
slim disk is shown in the right column. The upper panel shows
angular momentum in the disk (the solid line) in reference to the
Keplerian distribution (the dashed line). ISCO, indicated by the
dash-dotted line is at the radius where the Keplerian angular
momentum  has its minimum. The {\it potential spout} (a square) and the {\it
center} (a triangle) are defined as crossings of the angular
momentum in the disk line with the Keplerian line. For slim disks
they are at two different radii, on both sides of the ISCO. For
Shakura-Sunyaev disks they merge into one singular location at
ISCO. The lower panel shows the cross section of the disk. The
slim disk has everywhere a finite thickness while the
Shakura-Sunyaen disk is singular at ISCO (it has a zero thickness
there). The sonic radius (a cross) is where the accretion
component of the velocity equals the local sound speed. In slim
disks, the sonic point corresponds to a critical point of the set
of differential equations, that through the regularity conditions
defines the global eigensolution of the problem. The
Shakura-Sunyaev disk is described by local algebraic equations and
this global eigenvalue aspect is missing, thus location of a sonic
point is of no relevance.
%
%
\begin{acknowledgements}
This work was supported by Polish Ministry of Science grants N203
0093/1466, N203 304035, N203 380336, N203 00832/0709. AS
acknowledges support from the Department of Astronomy at Kyoto
University. MAA acknowledges a professorship at Universit{\'e}
Pierre et Marie Curie that supported his visit to Institut
d'Astrophysique in Paris during which a part of research reported
here was done. MAA also acknowledges the Czech government grant
MSM 4781305903. JPL acknowledges support from the French Space
Agency CNES.
\end{acknowledgements}


%
%
%
\begin{appendix}
%
%
\section{The Kerr geometry slim disks}
\label{sec:ss-vs-slim}
%
The Shakura-Sunyaev models are {\it local}: they are described by
algebraic equations, valid at any particular (radial) location in
the disk, independently of physical conditions at different
locations. Contrary to that, the slim disk models of accretion
disks are {\it non-local}. They are described by differential
equations globally connecting physical conditions at all radial
locations from infinity to the black hole horizon.

Initially, models of slim disks have been constructed by
\cite{abr-1988}, who used the pseudo-Newtonian potential of
\cite{wiita-1980} and Newtonian equations derived by
\cite{kogan-1981} and later improved by \cite{much-pacz-1982}, \cite{mat-1984} and
\cite{muchotrzeb-1983}. General relativistic version (the Kerr
metric) of the slim disk equations was derived and elaborated by
\cite{las-1994}, \cite{abr-1996},
\cite{gam-1998}, and most recently by \cite{sad-2009} who made
several corrections and improvements to results of the previous
authors, and who numerically constructed slim disk solutions in a
wide range of parameters applicable to the X-ray binaries. In
particular, he calculated the solutions in the whole relevant
range of accretion rates, from very sub-Eddingtonian, to
moderately super-Eddingtonian ones. In this paper we follow
notation and conventions adopted by \cite{sad-2009}. The Kerr
geometry slim disk equations adopted here are:

(i) The mass conservation:
\begin{equation}
\label{appendix-mass}
 \dot M=-2\pi \Sigma\Delta^{1/2}\frac{V}{\sqrt{1-V^2}}
\label{eq_cont2}
\end{equation}
where $\Sigma=\int_{-h}^{+h}\rho\,dz\approx2H\rho$ is disk surface
density and $V$ is the gas radial velocity as measured by an
observer at fixed $r$ who co-rotates with the fluid. Here
\begin{equation}
\label{appendix-Delta}
\Delta = r^2 - 2M\,r - a^2.
\end{equation}
\citep[For the Kerr metric description see e.g.][or any textbook
on general relativity]{kat-2008}. Equation (\ref{eq_cont2}) has
the same form in the Shakura-Sunyaev model.

(ii) The radial momentum conservation:
\begin{equation}
\frac{V}{1-V^2}\frac{dV}{dr}=\frac{\cal
A}{r}-\frac{1}{\Sigma}\frac{dP}{dr} \label{eq_rad3}
\end{equation}
where
\begin{equation}
\label{appendix-radial-momentum}
{\cal A}=
-\frac{MA}{r^3\Delta\Omega_k^+\Omega_k^-}\frac{(\Omega-\Omega_k^+)
(\Omega-\Omega_k^-)}{1-\tilde\Omega^2\tilde R^2} \label{eq_rad4}
\end{equation}
%
\begin{equation}
\label{appendix-A}
A = r^4 + r^2a^2 + 2M\,ra^2\,,
\end{equation}
$\Omega=u^\phi /u^t$ is the angular velocity with respect to the
stationary observer, $\tilde\Omega=\Omega-\omega$ is the angular
velocity with respect to the inertial observer, $\Omega_k^\pm=\pm
M^{1/2}/(r^{3/2}\pm aM^{1/2})$ are the angular frequencies of the
co-rotating and counter-rotating Keplerian orbits and $\tilde
R=A/(r^2\Delta^{1/2})$ is the radius of gyration. In the
Shakura-Sunyaev model this equation is a trivial identity $0=0$
because the radial pressure and velocity gradients vanish, and
rotation is Keplerian, $\Omega = \Omega_k^+$.

(iii) The angular momentum conservation:
\begin{equation}
\label{appendix-angular-momentum}
 \frac{\dot{M}}{2\pi}({\cal L}-{\cal L}_{in})=
 \frac{A^{1/2}\Delta^{1/2}\gamma}{r}\alpha P
\label{eq_ang6}
\end{equation}
where ${\cal L}=-u_\phi$ is the specific angular momentum, $\gamma$
is the Lorentz factor and $P=2Hp$ can be considered to be
vertically integrated pressure. The constant $\alpha$ is the
standard {\it alpha viscosity parameter} introduced by
\cite{sha-sun-1974}. The constant ${\cal L}_{in}$ is the angular
momentum at the horizon, unknown a priori. It provides an
eigenvalue linked to the unique
eigensolution to the set of slim disk differential equations
constrained by proper boundary and regularity conditions. The
algebraic equation (\ref{eq_ang6}) is the same as in the
Shakura-Sunyaev model, except that the Shakura-Sunyaev model {\it
assumes} that ${\cal L}_{in} = {\cal L}_k({\rm ISCO})$.

(iv) The vertical equilibrium:
\begin{equation}
\label{appendix-vertical}
 \frac{P}{\Sigma H^2}=\frac{{\cal L}^2-a^2(\epsilon^2-1)}{2 r^4}
\label{eq_vert}
\end{equation}
with $\epsilon=u_t$ being the conserved energy associated with the
time symmetry. The same equation is valid for the Shakura-Sunyaev
model.

(v) The energy conservation:
\be
\label{appendix-energy}
 -\frac{\alpha P
A\gamma^2}{r^3}\frac{d\Omega}{dr}-\frac{32}{3}\frac{\sigma
T^4}{\kappa\Sigma}= -\frac{\dot M}{2\pi
r\rho}\frac1{\Gamma_3-1}\left(\der pr-\Gamma_1\frac p\rho\der\rho
r\right)
\ee
here $T$ is the disk central temperature. The right hand side of
this equation represents the advective cooling and vanishes in the
Shakura-Sunyaev model. Because in the Shakura-Sunyaev model
rotation is Keplerian, $\Omega = \Omega_k^+$, which means that
$\Omega$ is a known function of $r$ and therefore the first term
on the left-hand side (which represents viscous heating) is
algebraic. The second term, which represents the radiative cooling
(in diffusive approximation) is similar in the
Shakura-Sunyaev model.
%
\section{No torque at the black hole horizon}
\label{appendix-horizon-torque}
%
The assumption about (vanishingly) small torque in the region
between black hole and accretion disk is well motivated
physically. Let us recall that the very meaning of a torque $Q$ is
that it transports angular momentum without transporting mass.
Correspondingly, the total angular momentum flux ${\dot J}$
through a surface equals, in general,
\be {\dot J} = {\dot M}j + Q,
\label{torque-definition}
\ee
where ${\dot M}$ is the mass flux, and $j$ is the angular momentum
density (per unit mass). However, the torque is only a
phenomenological concept. Microscopically, the flux ${\dot J}$
should be seen as a difference of material fluxes that come from
the opposite sides of the surface, ${\dot J} = [{\dot M}_+j_+] -
[{\dot M}_-j_-]$. One also has ${\dot M} = {\dot M}_+ - {\dot
M}_+$, and $j = (j_+ + j_-)/(j_+{\dot M}_+ + j_-{\dot M}_-)$.
Microscopically then, the torque is equal $Q = 2{\dot M}_+\,{\dot
M}_-(j_+ - j_-)/({\dot M}_+ + {\dot M}_-)$. It necessarily
vanishes when all matter crosses the surface in only one
direction, i.e. when either ${\dot M}_+ = 0$ or ${\dot M}_- = 0$.
As the only one-side matter flux is the fundamental property of
the black horizon, one concludes that there should be no torque at
the black hole surface.

Since the \cite{blandford-znajek} process energizes the jet (and
disk) by extracting rotational energy of a black hole through a
kind of electromagnetic braking, some astrophysicists argue that
in this case there must be a ``Maxwell'' torque between the black
hole and outside matter. However, by looking at the
Blandford-Znajek process from the quantum electrodynamics
perspective, one sees only ingoing, but not outgoing
photons. Thus, there is only one-way traffic of photons, and no
torque possible. The photons with {\it negative} energy and
angular momentum that are present in the ergosphere, are
responsible for the slowing down the hole, similarly to negative
energy particles in the classic Penrose process that must
necessarily have also a negative angular momentum. This point of
view, that the Blandford-Znajek process is an electromagnetic
version of the Penrose process, was recently discussed in context
of the classical Maxwell electrodynamics (in Kerr geometry) by
several authors, in particular most forcefully by \cite{kom-2008}.

Here, we generalize Komissarov's point to {\it any} material
field, not only the electromagnetic one. Following Komissarov, let
us consider the local ZAMO (or FIDO) observer in the Kerr
geometry. His four velocity in terms of the Killing vectors
$\eta^i$ (time symmetry) and $\xi^i$ (axial symmetry) is given by
$n^i = q(\eta^i + \omega\xi^i)$, where $\omega$ is the angular
velocity of frame dragging, and $q > 0$ follows from normalization
$n^in^k g_{ik} = -1$. Let us now consider a general matter or
field, described by an unspecified stress-energy tensor
$T^i_{~k}$. The energy flux in the ZAMO frame is $E^i = -
T^i_{~k}n^k$. The energy acquitted by the black hole is
\be
\label{komissarov-total-energy}
E = - \int T^i_{~k}n^k\,dN_i > 0,
\ee
where $\int\,dN_i$ is the surface integral over the horizon. The
inequality sign follows from the fact that the locally measured
energy must be positive. The above integral may by transformed
into
\be
\label{komissarov-condition}
0 < E = - \int q\,T^i_{~k}(\eta^k + \omega \xi^k)\,dN_i =
q_H(E_{\infty} - \omega_H J_{\infty}),
\ee
where the index $H$ denotes horizon, and $E_{\infty}$ and
$J_{\infty}$ are the ``energy at infinity'' and the ``angular
momentum at infinity'' acquired by the black hole absorbing the
corresponding fluxes of these quantities defined by,
\be
\label{komissarov-fluxes}
E^i_{\infty} = - T^i_{~k}\eta^k, ~~~J^i_{\infty} = T^i_{~k}\xi^k.
\ee
From (\ref{komissarov-condition}) one concludes  that $E_{\infty}
> J_{\infty}\omega_H$. As in the classic Penrose process, the
necessary condition for the extraction of energy at infinity is
that the energy (at infinity) absorbed by a black hole is
negative, $E_{\infty} < 0$. This is equivalent to
$J_{\infty}\omega_H < 0$. Thus, in a way fully analogous to the
line of arguments that is made discussing the Penrose process, one
may say that if energy at infinity increases because the black
hole absorbed a negative-at-infinity energy, then the black hole
must also slow down by absorbing matter or electromagnetic flux
with negative angular momentum.
%
\end{appendix}
%
%
\end{document}